\documentstyle[eqsecnum,aps]{revtex}

\def\be{\begin{equation}}
\def\ee{\end{equation}}
\begin{document}
\draft
\title{On the crosscorrelation between
Gravitational Wave Detectors for detecting
association with Gamma Ray Bursts }
\author{
G. Modestino$^1$, A. Moleti$^{2,3}$
\\
}
\address{
${}^{1)}$ Istituto Nazionale di Fisica Nucleare INFN, Frascati\\
${}^{2)}$ Dipartimento di Fisica, Universita' di Roma "Tor Vergata"\\
${}^{3)}$ INFN - Sezione di Roma2, Roma,Italy}
\date{\today}
\maketitle
\begin{abstract}
Crosscorrelation of the outputs of two Gravitational Wave (GW) detectors
has recently been proposed \cite{finn} as a method for detecting 
statistical association 
between GWs and Gamma Ray Bursts (GRBs). Unfortunately, the method can be 
effectively used only in 
the case of stationary noise. In this work a different crosscorrelation 
algorithm is presented, which 
may effectively be applied also in non-stationary conditions for the cumulative analysis of a large 
number of GRBs. The value of the crosscorrelation at zero delay, 
which is the only one expected to be 
correlated to any astrophysical signal, is compared with the distribution 
of crosscorrelation of 
the same data for all non-zero delays within the integration time interval. 
This background 
distribution is gaussian, so the statistical significance of an 
experimentally observed excess would 
be well-defined. Computer simulations using real noise data of the cryogenic 
GW detectors Explorer and 
Nautilus with superimposed delta-like signals were performed, to test 
the effectiveness of the 
method, and theoretical estimates of its sensitivity compared to 
the results of the simulation. 
The effectiveness of the proposed algorithm is compared to that of other 
cumulative techniques, 
finding that the algorithm is particularly effective in the case of 
non-gaussian noise and of a large 
(100-1000s) and unpredictable delay between GWs and GRBs.
\end{abstract}

\section{Introduction}
Over the last decade, Gamma Ray Bursts have been successfully investigated
with the satellite experiments BATSE \cite{batse} and Beppo-SAX
\cite{costa}. The large database 
now available includes information, for more than 2,000 GRBs, on
the GRB arrival time, duration, intensity in some frequency bands, sky
position of the source, and (for a small GRB subset) redshift. On the
basis of all this
information it is possible to design cumulative algorithms to 
detect a statistical association 
between the Gravitational Wave (GW) detector signals and the GRBs.
It is well known that the present sensitivity of GW detectors is not
sufficient to detect single events unambiguously, with the exception of
very nearby GW sources, which are expected to be very rare. For this
reason, much effort has been devoted to the development of data analysis
techniques for the detection of coincidences between the events recorded
by different detectors. The main problem facing this kind of analysis is
that the event lists of both detectors are dominated by the contribution
of non-gaussian and non-stationary noise \cite{astone00}.
Observation of a large number of GRBs, probably associated
with explosive events capable of producing a large GW signal, has afforded
the possibility to analyze the GW detector data around the GRB
arrival times. Cumulative techniques have been proposed to
detect a statistically significant association between GW signals and
GRBs \cite{bonazzola,modestino97,modestino98,murphy,modestino00}.
A difficulty arises from the theoretical uncertainty about the time
delay between the GRB and GW arrival times. This delay depends on
the models used to describe GRB dynamics, and, however, it is not
expected to be constant. Theoretical predictions
\cite{piran,rees,meszaros},
and the interpretation of experimental results \cite{frontera}
based on a fireball model \cite{sari}
suggest that delays up to 1hr should be expected.
Thus, any cumulative analysis technique of GW detector data, 
synchronized to the GRB arrival time, 
would dramatically lose effectiveness due to the uncertainty about the 
GW-GRB time delay.
Recently, a crosscorrelation method
has been proposed by Finn et al. \cite{finn}, to detect statistical
association between GWs and GRBs. The method is based on measurement of
the average of the crosscorrelation of two detectors (the two LIGO
interferometers were considered), on a set (named on-GRB set) of
time windows centered at
the arrival times of the GRB events. An off-GRB set is
used as a reference, and the statistical significance of the difference
between the average crosscorrelation of the two sets is evaluated. 
As correctly claimed by the 
authors, the non-gaussian nature of the noise does not affect the
method, but the hypothesis of stationarity of the GW detector noise is
necessary to obtain a meaningful result. Gravitational wave detectors are 
typically affected by noise 
that is not only non-gaussian, but also non-stationary. The
non-stationarity of the noise and uncertainty about the GW-GRB time
delay make it difficult to choose a suitable off-GRB data set providing 
an unbiased background. In 
fact, it would be necessary to choose the off-GRB
data samples near those of the on-GRB set in order to decrease the
uncertainty introduced by the noise non-stationarity. On the other hand,
due to GW-GRB delay uncertainty, unbiased reference samples should be 
chosen far from the GRB trigger times.
Thus it is very important to design a different crosscorrelation 
technique, capable of overcoming this 
difficulty with optimal choice of the reference quantity. 
In this work it will be shown that the reasonably obvious assumption of
simultaneity of any real astrophysical GW signal on the 
two detectors may be used to design a 
simple technique for detecting association of GW signals with GRBs, 
which is affected neither by the non-stationarity nor the non-gaussianity of
the detectors' noise. Simultaneity is assumed to hold within
a Wiener filter characteristic time, which, for the narrow-band resonant GW
detectors considered here, is always much longer than the 
physical delay
associated with the distance between the GW detectors.
In Section 2 the proposed method will be described. The results of
numerical simulations and analytical computations will be shown and discussed 
in Section 3.

\section{Method}
Let $x_{i}(t)$ and $y_{i}(t)$ be the Wiener filtered outputs of 
the GW detectors X and Y, during a time 
interval T, centered at $t_{{\gamma}{i}}$, the time
of arrival of the i-th GRB. The data are placed on a circular buffer, 
to compute on the same data set 
the crosscorrelation between the two GW detectors as a 
function of delay $\tau$:
\begin{equation}
\chi_{i}(\tau) = \frac{1}{T}\int_{t_{{\gamma}{i}}-T/2}^{t_{{\gamma}{i}}+T/2}
dt x_{i}(t)y_{i}(t+\tau)
\label{croc}
\end{equation}
The average $<\chi_{i}>$ and the standard deviation
$\sigma_{{\chi}{i}}$ of the i-th $\chi_{i}(\tau)$
distribution are then computed, and the normalized, 
zero-mean quantity is obtained:
\begin{equation}
c_{i}(\tau) = \frac{\chi_{i}(\tau)-<\chi_{i}>}{\sigma_{{\chi}{i}}}
\label{citau}
\end{equation}
The functions $c_{i}(\tau)$ associated with 
each GRB are then averaged, yielding a 
cumulative zero-mean crosscorrelation, which is a function of delay $\tau$:
\begin{equation}
C(\tau) = \frac{1}{N} \sum_{i} c_{i}(\tau)
\label{c_ave}
\end{equation}
For sufficiently long integration times, the quantity $C(\tau)$ 
is a random variable with gaussian distribution. 
The crosscorrelation signal to 
noise ratio SNRC is then obtained by
dividing the zero-delay crosscorrelation C(0) by the standard deviation
$\sigma_{C}$ of the $C(\tau)$ distribution:
\begin{equation}
SNRC = \frac {C(0)}{\sigma_{C}}
\label{snrc}
\end{equation}
We note here that this procedure, truncated at the step of
 Eq.\ref{citau}, may also be applied to single GRB events.
The behavior of  a $c_{i}(\tau)$, and the
crosscorrelation snr, $SNRc_{i}$ may be evaluated for any single GRB.
Analysis of an individual GRB could be interesting in the case of very
peculiar GRBs (high power and/or low redshift). However, the method becomes 
particularly effective for 
the cumulative analysis of a large number of GRBs, 
because the cumulative SNR advantage, as the square 
root of the number of samples, is fully obtained also if the delay 
between the GRB and the GW is unknown and variable. 
The only requirement of the GRB-GW delay is that it must be smaller than the 
integration time, for all GRBs. In this case the crosscorrelation 
contributions due to differently 
delayed GW signals add coherently, while, in the case of other cumulative 
algorithms using one 
detector only, GRB-GW simultaneity is needed to obtain the full 
cumulative SNR advantage.
It should be added that the integration interval doesn't need to be centered 
around the GRB arrival time, as it is in Eq.\ref{croc}. It may be 
arbitrarily shifted to test the predictions of different theoretical models. 

\section{Results and Discussion}
\begin{figure}[t]
\vspace*{8.0cm}
\centering
\includegraphics{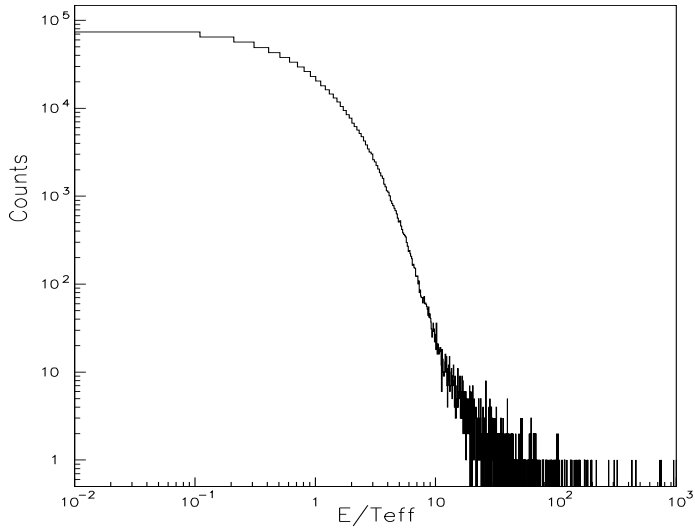}
\caption{Energy histogram of the Wiener filtered data of the detectors in
the 27 selected common hours. The energy is normalized to $T_{eff}$ of the
selected data.
\label{distri1}}
\end{figure}

The sensitivity of the above-described procedure has been evaluated, both
analytically and with computer simulations,
using real Wiener filtered \cite{astone94} data of the
GW detectors Nautilus \cite{astone97} and
Explorer \cite{astone93}.
A period of one year of GW data was considered. The data
sampling time is 0.296 s. The adaptive Wiener filter smooths the noise with the
time constant $\tau_{3}\approx{1}$ s  for both detectors, corresponding to an 
effective bandwidth $\beta_{3}\approx{1}$ Hz. The noise level of the filtered 
data is expressed by the effective temperature $T_{eff}$, which is the minimum 
energy variation detectable by the antenna with signal to noise ratio 
equal to unity.
One-hour-long common data stretches were
used, centered at the times given by a dummy GRB time list with the 
same experimental BATSE GRB rate (1/day) \cite{batse}. 
The data stretch associated with each GRB was selected for the 
analysis if the
average effective temperature proved less than 20 mK on both detectors. This 
choice was suggested by the noise distribution of the detectors in the 
considered year. The noise distributions of both detectors were peaked in 
the 10-15 mK range, so a higher threshold would have not significantly 
improved the statistics.
The above constraints yielded N = 27 selected one-hour common data stretches
of 12000 samples each, with an average effective temperature $T_{eff}$ = 13
mK. These figures give the important information about the size of the 
statistical sample that can 
reasonably be obtained for a crosscorrelation analysis of two real GW 
detectors. It is clear that the 
requirement of having both detectors simultaneously in operation 
with low noise performance severely 
reduces the size of the available statistical sample. The same constraint 
applied to one detector only 
would give a much larger sample (N=150). This is clearly a drawback of 
the proposed method, which 
could be limited by using pairs of data coming from more than two GW detectors.
In Fig.\ref{distri1} the energy histogram of  the filtered data of both 
detectors is
shown. The energy is normalized to $T_{eff}$ of the selected data. Looking at
the distribution of Fig.\ref{distri1}, it is  clear that the large 
non-gaussian tail
of the energy event distribution makes it impossible to
detect unambiguously individual events with $E/T_{eff}$ = 5-10. 

\begin{figure}[t]
\vspace*{8.0cm}
\centering
\includegraphics{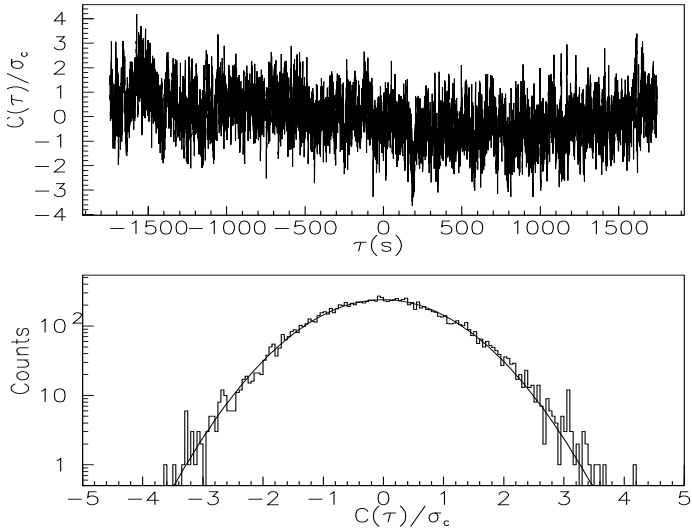}
\caption{Cumulative crosscorrelation of the noise of the two GW detectors, and
its statistical distribution, which is clearly well approximated 
by a gaussian function.
\label{distri2}}
\end{figure}

This is a well-known
problem in GW data analysis, which has generally led the GW data
analysis community to define as GW events, to be used for a coincidence
analysis, those with a very large value of $E/T_{eff}$.
The procedure was applied to the N one-hour periods. The choice of
using one hour integration time is proposed to include the contribution
of GW signals associated to GRBs, according to the delay predicted by
most theoretical models. This choice could be optimized, as will be 
discussed later.
In Fig.\ref{distri2} the normalized average crosscorrelation
$C(\tau)/\sigma_{C}$ and its histogram are shown for the 27 selected hours. 
The distribution of the crosscorrelation variable
$C(\tau)/\sigma_{C}$  shows a gaussian shape, 
without any significant tail, while the original 
distributions of the two detectors' noise were both markedly non-gaussian. 

\begin{figure}[t]
\vspace*{8.cm}
\centering
\includegraphics{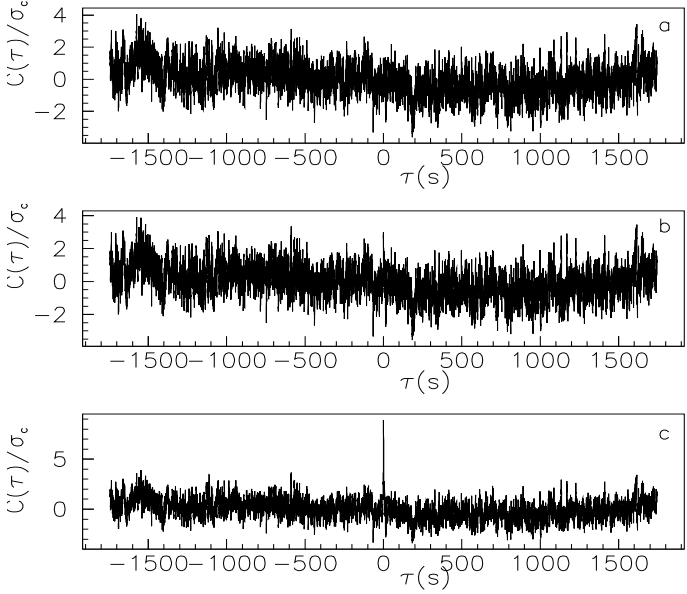}
\caption{Cumulative crosscorrelation of the noise of the two GW detectors, 
Nautilus and Explorer,
with superimposed signals of increasing energy: 2.2$T_{eff}$, 4.5$T_{eff}$. 9.1$T_{eff}$. The 
superimposed GW signals are simultaneous on the two GW detectors, but variably delayed with respect to 
the GRB arrival time.
\label{distri3}}
\end{figure}

This result, in the light of the central 
limit theorem, is a quite obvious consequence of the averaging processes 
(time integration and sample 
average) applied to the original data to give the variable $C(\tau)$. 
The gaussianity of the 
crosscorrelation variable is very important, as it gives a 
well-defined statistical meaning to the 
experimental result. The probability that a given high value of the 
experimental crosscorrelation 
quantity be due to chance may reliably be computed, as well as the 
upper limit on the source average 
power implied by a low experimental value.\\
An analytical estimate of the sensitivity of the method has been obtained, 
assuming gaussian noise. The 
expected normalized zero-mean crosscorrelation SNRC is given by:
\begin{equation}
SNRC = ({SNR_{x} SNR_{y} +SNR_{x}+SNR_{y}})\sqrt{\frac {N}{N'}}
\label{snrct}
\end{equation}
where $SNR_{x}$ and $SNR_{y}$ are the snr's of the two detectors, in terms 
of E/$T_{eff}$, and N' = T 
$\beta_{3}$, is the number of independent samples in the integration interval.
A numerical simulation was performed to test the sensitivity of the algorithm using the non gaussian 
noise data of two real GW detectors, Nautilus and Explorer.
Impulsive signals of energy E were added to the N stretches of noise data, 
randomly delayed with respect to $t_{{\gamma}{i}}$ within the integration 
interval, but simultaneous on the two detectors X and Y.
In Fig.\ref{distri3} the normalized cumulative crosscorrelation $C(\tau)/\sigma_{C}$ is plotted as a 
function of $\tau$, for three values of the added signal amplitude, corresponding to increasing SNR on 
the single detector:
$E/T_{eff}$ = 2.2, 4.5, 9.1. The data of Fig.\ref{distri3} show
the corresponding crosscorrelation signal C(0), emerging from the gaussian
background. Low $E/T_{eff}$ (e.g. between 3 and 10) events, which would be
totally immersed in the non-gaussian tails of Fig.\ref{distri1}, distinctly
emerge from the gaussian tails of the distribution of Fig.\ref{distri2}.\\

\begin{figure}[t]
\vspace*{6.cm}
\centering
\includegraphics{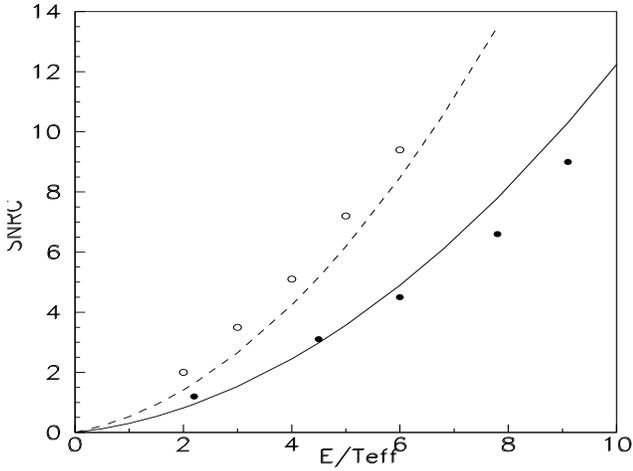}
\caption{ Crosscorrelation snr, SNRC,  plotted as a function of the added
signal snr, $E/T_{eff}$ ($\bullet$ for T = 1 hr,  $\circ$ for T = 20 min).
The numerical results are compared to the analytical estimation 
Eq.\ref{snrct}.
\label{snr4}}
\end{figure}

The sensitivity of the algorithm, as computed by the numerical simulations, 
is shown in Fig.\ref{snr4}, 
where the adimensional crosscorrelation signal to noise ratio SNRC, defined by 
Eq.\ref{snrc}, is 
plotted as a function of the signal SNR,
$E/T_{eff}$, for two values of the integration time T. The 
simulation shows that SNRC decreases as the 
square root of the integration time, provided that T 
is larger than the maximum delay between GRBs and 
GW signals, as in this simulation, and that the relation between 
SNRC and $E/T_{eff}$ is in agreement with Eq.\ref{snrct}.
As discussed above, a given value of the SNR is much more significant for the
crosscorrelation variable, due to the gaussianity of the crosscorrelation
background.
The sensitivity of the algorithm may be compared to that of a 
low-threshold coincidence search, which 
is a possible alternative two-detector method.
Setting an event threshold at $SNR_{t}$ one can compute \cite{pizze} the 
probability of getting an event larger than the threshold in 
the presence of gaussian noise, as a 
function of the signal SNR, $SNR_{s}=E/T_{eff}$:

\begin{equation}
P_{x}(SNR_{s},SNR_{t}) = \int_{SNR_{t}}^{\infty}
dx\frac{\exp{\frac{-(SNR_{s}+x)}{2}}\cosh{\sqrt{xSNR_{s}}}}{\sqrt{2{\pi}x}}
\label{psnr}
\end{equation}

The expected number of coincidences due to the N added signals would be:

\begin{equation}
N_{C} = P_{x}P_{y}N
\end{equation}
For $SNR_{s}=SNR_{t}$ we obtain $P_{x}=P_{y}=0.5$ and the coincidence
excess is $N_{C}\approx{7}$.
The coincidence background during the whole time interval NT 
may also be computed for gaussian noise:

\begin{equation}
N_{acc} = P(0,SNR_{t}) P(0,SNR_{t}) N N'
\end{equation}

The coincidence excess with respect to the Poisson distribution of the 
accidental coincidences has a probability of being due to chance 
that may be compared to the corresponding 
probability of getting by chance the crosscorrelation excess with 
respect to the gaussian crosscorrelation background found with the simulation. 
This comparison is shown in Table \ref{compa} for three 
values of the coincidence SNR threshold, in the case $SNR_{s} = SNR_{t}$
and $T=1h$. $P_{SNRC}$ is the probability of finding by chance the 
corresponding value of SNRC, assuming gaussian statistics. $P_{c,g}$
is the probability of finding by chance the coincidence excess 
$N_{C}$, assuming the accidental coincidences background $N_{acc,g}$, 
computed in the case of gaussian noise on both detectors.$P_{c,r}$ 
is the probability of finding by chance the same coincidence excess 
$N_{C}$, assuming the accidental coincidences background $N_{acc,r}$, 
computed from the real event distribution of 
the detectors, shown in Fig.\ref{distri1}.
The evaluation of the accidental coincidences background based on gaussian 
statistics is not adequate to describe real GW detectors, whose noise 
is not gaussian. A more realistic evaluation of the accidental 
coincidences background is 
found using the real event rate of the two GW 
detectors as a function of the threshold level $SNR_{t}$. 
From Table \ref{compa} it is clear that the coincidence technique 
would be more effective 
for a hypothetical detector with gaussian noise, while the crosscorrelation 
method proposed here proves preferable in the more realistic case of non 
gaussian noise. It should also be added that the coincidence search method 
is also affected, for non-delta-like signals, by an uncertainty about the event 
maximum time that leads either to a decrease of the coincidence detection 
efficiency, or to the choice of a longer coincidence window, with a 
correspondingly higher rate of accidental coincidences.
Of course, other cumulative single-detector techniques could also be 
effective, but only in the case 
of close simultaneity between the GW event and the GRB
trigger time. This model-dependent assumption is not needed for the
effectiveness of the proposed crosscorrelation technique.
The advantage of the proposed method is the absence of hypotheses,
excluding the obvious assumptions that the signals of the two GW
detectors be simultaneous. No hypothesis is made on both the gaussianity and
the stationarity of the noise. A similar method, proposed by Finn et al.
\cite{finn}, is critically based on the hypothesis of 
stationarity of the GW detector
noise, which is not generally true for present GW detectors. The intrinsic 
fluctuation of the noise contribution to the crosscorrelation of the on-GRB 
sample would require a very large number of low-noise data samples. 
The advantage of the method proposed here has nothing to do with the 
extraction of a 
crosscorrelation quantity with the maximum signal contribution. 
Rather, it lies in the choice of the reference quantity to which the
measured crosscorrelation is compared to evaluate its statistical
significance. In the case of Finn et al., the reference quantity is
found by choosing reference data uncorrelated to the GRB arrival times, 
according to the noise stationarity hypothesis. In the present work, 
only the obvious physical hypothesis that the signals be 
simultaneous on the two GW detectors is used to define the reference 
quantity, which makes it possible to
avoid the problem of non-stationarity of the GW detector noise.
Model-dependent assumptions on the GRB physics could be considered to
increase the sensitivity of the method. For example, the uncertainty
about the time delay between GRBs and GWs would suggest computing the
crosscorrelation on a time window wide enough to include this
uncertainty. As the sensitivity of the method is decreasing with the
width of the window, the model-dependent hypothesis that the delay uncertainty
is correlated to the GRB duration could be used, for
example, to optimize the method by choosing each time window width
according to the GRB duration. 
Unfortunately, as pointed out above, the obvious drawback of 
any crosscorrelation 
technique is the reduced size of the data sample, due to the requirement of 
simultaneous low-noise operation of two GW detectors.
Multiple crosscorrelation with n detectors
is also a natural extension of the proposed method.

\begin{table}
\caption{Comparison between the crosscorrelation and the low-threshold 
coincidence technique.}
\label{compa}
\begin{tabular}{ccccccc}
{$E/T_{eff}$}&{$SNRC(T=1h)$}&{$P_{SNRC}$}&{$N_{acc,g}$}&
{$P_{c,g}$}&{$N_{acc,r}$}&$P_{c,r}$ \\
\tableline
{$2$}&{$0.8$}&{$0.21$}&{$200$}&{$0.3$}&{$2000$}&{$0.4$}\\
{$3$}&{$1.5$}&{$7\cdot10^{-2}$}&{$1$}&{$10^{-5}$}&{$250$}&{$0.3$}\\
{$5$}&{$3.5$}&{$2\cdot10^{-4}$}&{$0$}&{}&{$22$}&{$9\cdot10^{-2}$}\\
\end{tabular}
\end{table}

\section{Conclusions}
A new method is proposed to detect statistical association between 
GW detector signals and GRBs, by 
using a cumulative crosscorrelation technique. The originality of the 
proposed method lies in the choice of 
the reference background quantity to which the crosscorrelation should be 
compared, thanks to which unbiased meaningful results can also be obtained 
in the case of non-stationary noise. 
The obvious physical 
constraint, i.e. that the signals must be simultaneous on the two GW detectors, 
is used to select as 
physically relevant the crosscorrelation at zero delay only, while the 
crosscorrelation integrals computed on 
the same circularly permuted data for other delays are used as a background 
distribution, providing a 
well-defined estimate of the statistical significance 
of the zero-delay result, in the eventuality of an excess, 
or of an absence of excess.
Computer simulations using the real noise data of the cryogenic GW detectors 
Nautilus and Explorer 
have been performed and compared to analytical estimates of the algorithm 
sensitivity. The sensitivity of the proposed algorithm has also been compared 
to that of a low-threshold coincidence search method, finding that the 
crosscorrelation method proves more effective in the case of non-gaussian 
noise.\\
\\
{\textit{\textbf{Acknowledgments}}\\
We thank F.Frontera, G.Pizzella, P.Bonifazi, 
G.V.Pallottino and E.Coccia for useful contributions, and 
the colleagues of the ROG group for providing the Explorer 
and Nautilus noise data used to test the 
algorithm effectiveness.\\

\end{document}